\documentclass[review]{elsarticle}
\usepackage[latin9]{inputenc}
\usepackage{url}
\usepackage{graphicx}

\makeatletter

\providecommand{\tabularnewline}{\\}

\usepackage{hyperref}
\usepackage{lineno}
\modulolinenumbers[5]

\journal{Computers \& Security}

\@ifundefined{showcaptionsetup}{}{%
 \PassOptionsToPackage{caption=false}{subfig}}
\usepackage{subfig}
\makeatother

\begin{document}
\begin{frontmatter}

\title{Robust Keystroke Biometric Anomaly Detection}

\author{John V.~Monaco\corref{correspondingauthor}}

\address{U.S.~Army Research Laboratory\\
Aberdeen Proving Ground, MD 21005, USA}

\cortext[correspondingauthor]{Corresponding author}

\ead{john.v.monaco2.civ@mail.mil}

\ead[url]{www.vmonaco.com}
\begin{abstract}
The Keystroke Biometrics Ongoing Competition (KBOC) presented an anomaly
detection challenge with a public keystroke dataset containing a large
number of subjects and real-world aspects. Over 300 subjects typed
case-insensitive repetitions of their first and last name, and as
a result, keystroke sequences could vary in length and order depending
on the usage of modifier keys. To deal with this, a keystroke alignment
preprocessing algorithm was developed to establish a semantic correspondence
between keystrokes in mismatched sequences. The method is robust in
the sense that query keystroke sequences need only approximately match
a target sequence, and alignment is agnostic to the particular anomaly
detector used. This paper describes the fifteen best-performing anomaly
detection systems submitted to the KBOC, which ranged from auto-encoding
neural networks to ensemble methods. Manhattan distance achieved the
lowest equal error rate of 5.32\%, while all fifteen systems performed
better than any other submission. Performance gains are shown to be
due in large part not to the particular anomaly detector, but to preprocessing
and score normalization techniques.\end{abstract}
\end{frontmatter}

\section{Introduction}

The Keystroke Biometrics Ongoing Competition (KBOC) presented an anomaly
detection challenge in the form of a biometric competition \citep{morales2016kboc,morales2016keystroke}.
Given only several template samples for training, participants were
tasked with designing keystroke biometric verification systems that
achieved a low equal error rate on a set of unlabeled query samples.
Unlike previous keystroke biometric competitions \citep{monaco2015one},
the KBOC utilized a platform that allows ongoing participation \citep{morales2015keystroke}.
This paper describes the best-performing systems of the KBOC. These
systems addressed two unique features of the KBOC dataset not typically
seen in other keystroke datasets.

\emph{The keystrokes closely match, but are not necessary identical
to, a target sequence}. Genuine and impostor subjects were required
to type a \emph{case insensitive} match to the template subject's
first and last name. As a result, the samples contain insertions,
deletions, substitutions, and transpositions from each other, primarily
due to the usage of modifier keys. Even the template keystroke sequences
sometimes differ from one another, reflecting inconsistencies in the
way a genuine subject types their name. Thus, the template and query
samples only approximately match a target keystroke sequence. This
scenario is not quite fixed-text, such as password entry, and not
quite free-text, such as an essay response. To deal with the mismatched
samples, a keystroke alignment preprocessing algorithm robust to typing
errors was developed. The method maximizes the amount of data available
for training and testing, achieving better performance over benchmark
alternatives, such as discarding mismatched keystrokes or truncating
sequences to the length of the shortest sequence.

\emph{The target keystroke sequence is unique to each subject}. Unlike
most fixed-text datasets in which every subject types the same password
or passphrase (e.g., \citep{killourhy2009comparing} and other datasets
described in \citet{giot2015review}), the target input string\footnote{\emph{string} refers to the character sequence that actually appears
on screen, and \emph{keystroke sequence} refers to the sequence of
keys pressed. For example, both keystroke sequences ``\texttt{G},
\texttt{Backspace}, \texttt{T}'' and ``\texttt{T}'' result in string
``t''. Keyboard key names are denoted in \texttt{monospace}.} for each subject in the KBOC dataset was the subject's first and
last name. Thus, the keystroke sequences for each subject are different.
Importantly, the samples reflect a keystroke sequence that the genuine
subject is more familiar with than an impostor. Since the keystroke
sequences are different for each subject, two-class models (i.e.,
genuine vs.~impostor) are not practical due to the difficulty in
utilizing negative data for training. Therefore, only one-class models,
i.e, anomaly detectors, are considered. Given only several template
samples are available for training, the anomaly detector must be able
to learn a given subject's pattern quickly. To this end, feature and
score normalization techniques robust to outliers and few training
samples were developed.

The keystroke alignment and score normalization techniques described
in this work are agnostic to the particular anomaly detector used,
and results are obtained for a range of models. The complete anomaly
detection systems utilize a processing pipeline that consists of four
main steps: keystroke alignment, feature extraction, scoring, and
score normalization. During keystroke alignment, a correspondence
is established between keystrokes of the query and template samples,
which may differ slightly as described below. The query sample keystrokes
are aligned to match the template, which enables standard fixed-text
techniques to subsequently be applied. Keystroke timing features are
extracted from the aligned sequences using normalization bounds specific
to each subject. Following this, an anomaly detector is trained on
the template samples and a similarity score is assigned to the query
sample. Finally, the score is normalized using subject-specific normalization
bounds.

Performance gains are largely due to the score normalization and keystroke
alignment techniques, over alternative methods, as demonstrated in
the results of this work. The feature and score normalization techniques
use subject-specific bounds which are robust against feature and score
outliers, respectively. The keystroke alignment method is robust to
typing errors and slightly differing keystroke sequences. This maximizes
the amount of data available for training and testing, leading to
a reduced failure to capture (FTC) rate \citep{bours2014performance}.
The fifteen best-performing systems in the KBOC ranged from auto-encoding
neural networks to Manhattan distance, and all fifteen systems achieved
a lower equal error rate than any other submission.

The rest of this paper is organized as follows. Section \ref{sec:Background}
provides a background of the KBOC data collection procedure and competition
details. Sections \ref{sec:Keystroke-alignment} and \ref{sec:Feature-extraction}
describe keystroke alignment and feature extraction, respectively.
Section \ref{sec:Anomaly-detection} describes the various anomaly
detectors, followed by score normalization in Section \ref{sec:Score-normalization}.
Experimental results are contained in Section \ref{sec:Results},
and Section \ref{sec:Discussion} discusses and analyzes the effects
of keystroke alignment and score normalization. Finally, Section \ref{sec:Conclusions}
draws conclusions and discusses future work. Source code of all the
systems described and to reproduce the results in this paper is available
at \url{https://github.com/vmonaco/kboc}.

\section{Background\label{sec:Background}}

The KBOC was part of the \emph{IEEE Eighth International Conference
on Biometrics: Theory, Applications, and Systems} (BTAS 2016) and
had two modes of participation: online mode and offline mode. Participation
in the online mode is continuous, hence the ``ongoing'' competition\footnote{Current leaderboard \url{https://www.beat-eu.org/platform/search/aythamimm/KBOC16_COMPETITION_SEARCH/}}
\citep{morales2016kboc,morales2016keystroke}. This paper describes
only the offline mode, which ran from January 31, 2016 (when the test
dataset became available), to April 22, 2016 (the deadline for submissions).
Each participant could submit up to 15 systems, and submissions were
evaluated only after the competition deadline to avoid overfitting.
The KBOC dataset, described in this section, is publicly available
upon request\footnote{See \url{https://sites.google.com/site/btas16kboc/home}}.

\subsection{Dataset}

Two separate datasets were made available to competition participants:
a development set for the purpose of system development, and a test
set for the purpose of system evaluation. The development set contains
10 subjects, each with 24 labeled samples (14 genuine and 10 impostor).
The test set contains 300 subjects, each with 4 labeled template samples
and 20 unlabeled query samples. The ratio of genuine to impostor samples
in the test set remained hidden from competition participants to increase
difficulty.

Data collection was performed in a semi-controlled environment designed
for multimodal recordings \citep{fierrez2010biosecurid}. Subjects
were geographically dispersed across 6 collection sites and approximately
evenly distributed by age (42\% 18\textendash 25, 22\% 25\textendash 35,
16\% 35\textendash 45, 20\% $>$45) and gender (54\% male). Each subject
participated in 4 data collection sessions within 4 months time. During
each session, the subject typed 4 case-insensitive repetitions of
their own name (2 in the middle of the session and 2 at the end) and
the names of 3 other subjects (7 samples total). While there were
totally 16 genuine and 12 impostor samples available for each subject
in the original dataset \citep{fierrez2010biosecurid}, the numbers
of genuine and impostor samples were randomly selected for the KBOC
test dataset. This ranged from 8 to 12, leaving 4 template samples
per subject. Samples that contained typing errors during data collection
were discarded. However, the number of keystrokes per subject could
vary due to the use of modifier keys, such as \texttt{Shift} and \texttt{Caps
Lock}. This is an important characteristic motivating the development
of a keystroke alignment preprocessing algorithm in Section \ref{sec:Keystroke-alignment}.

Each raw sample contains a sequence of 3-tuple events comprised of
the action (press or release), keyboard scancode, and the time interval
since the previous event (beginning with 0 for the first event). In
this work, the action sequence is converted to a keystroke sequence,
wherein each keystroke event is a 3-tuple that contains the key name,
press timestamp, and release timestamp. Samples range in length based
on the subject's name, from 12 to 30 keystrokes with an average of
$25.5\pm4.4$ keystrokes. Timing features, described in Section \ref{sec:Feature-extraction},
are extracted from the aligned keystroke sequence.

\begin{figure}
\begin{centering}
\subfloat[KBOC keystroke timings with estimated 46.4 ms resolution.\label{fig:KBOC-keystroke-latency}]{\begin{centering}
\includegraphics[width=1\textwidth]{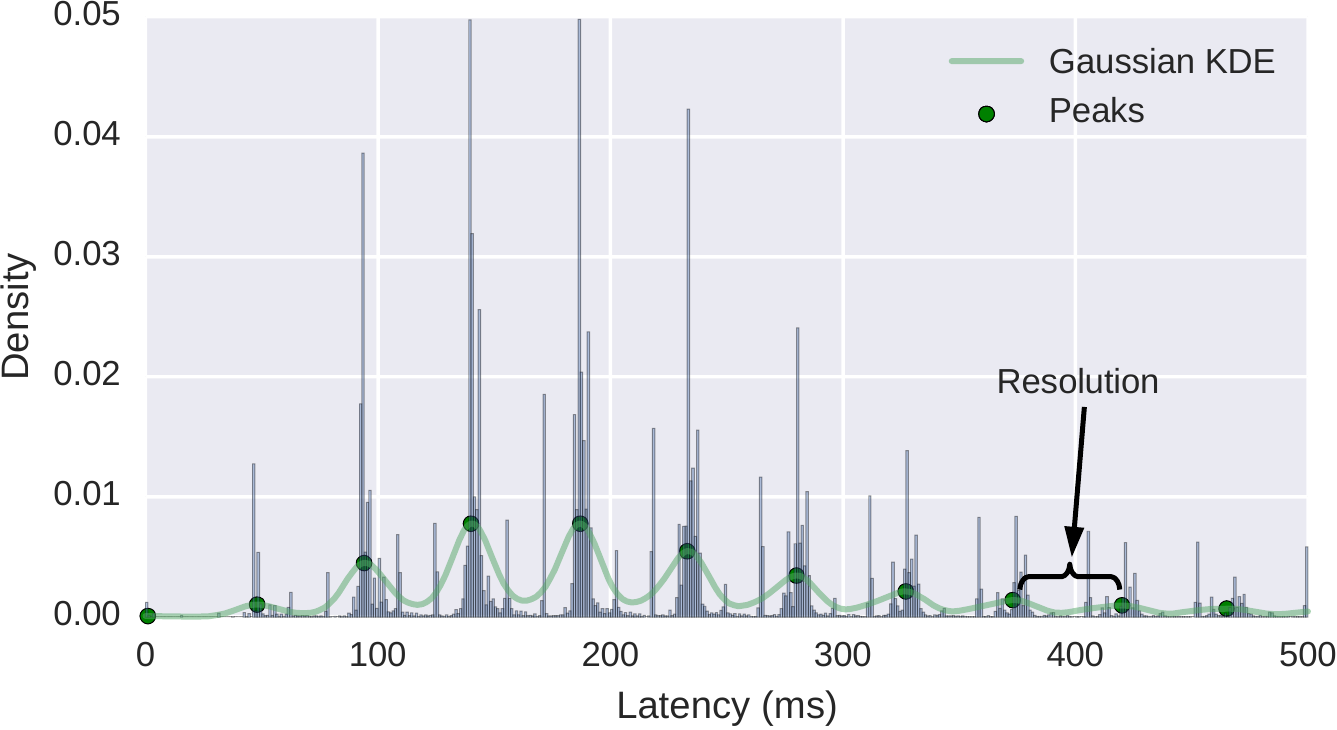}
\par\end{centering}

}
\par\end{centering}

\begin{centering}
\subfloat[\citet{monaco2013recent} keystroke timings with estimated 15.6 ms
resolution.\label{fig:Villani-keystroke-latency}]{\begin{centering}
\includegraphics[width=1\textwidth]{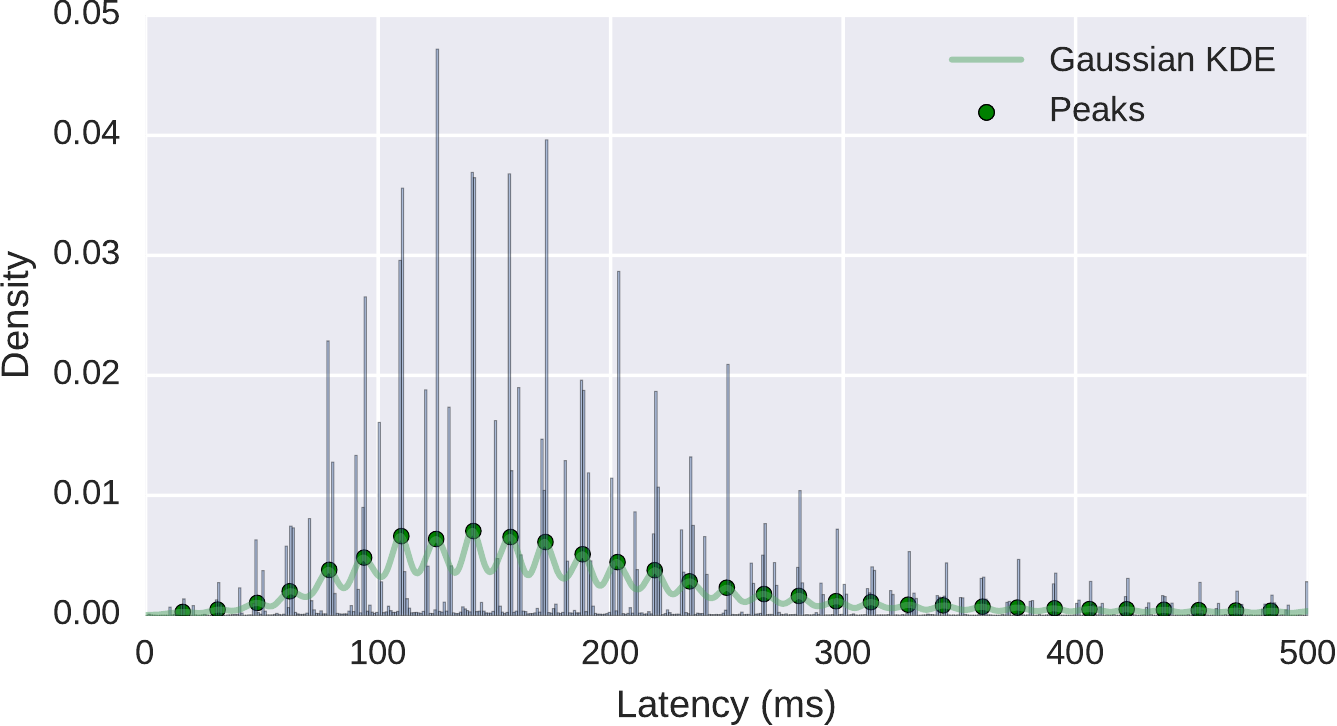}
\par\end{centering}

}
\par\end{centering}

\caption{Resolution of the KBOC keystroke timings compared to another dataset.
Resolution is estimated by taking the difference between modes of
a Gaussian KDE applied to the empirical press-press latency distribution.\label{fig:Precision}}
\end{figure}

Timestamps were recorded with 1 millisecond (ms) resolution, although
the actual resolution is limited by scheduling policies of the operating
system kernel. Specifically, most operating system kernels handle
interrupts from peripheral devices on a fixed period set by a global
timer. Timestamp resolution is then limited by how quickly the kernel
can respond to an interrupt generated by the keyboard device, which
is at most the global timer period. This effect is evident in the
empirical distribution of press-press latencies, shown in Figure \ref{fig:KBOC-keystroke-latency}.
A Gaussian kernel density estimate (KDE) is applied to the latencies
to identify regions of relatively high density, i.e., clustered latencies.
The high density regions reflect quantization introduced by the global
timer, and dispersion within each region reflects timestamp precision\footnote{Note that resolution is the degree to which a measurement can be made
and precision is the degree to which a measurement can be repeated.}. Taking the mean distance between modes of the Gaussian KDE gives
an estimated precision of 46.4 ms, close to the 40 ms estimated precision
in \citet{morales2016keystroke}. For comparison, the same technique
is applied to the press-press latencies of a free-text keystroke dataset
collected on a different platform: a Java application running on Windows
desktop computers \citep{monaco2013recent}. The resolution of the
free-text dataset, shown in Figure \ref{fig:Villani-keystroke-latency},
is estimated to be 15.6 ms, which agrees very well with the 15.6 ms
Windows default system-wide timer \citep{windowstimer}. According
to results in \citet{killourhy2008effect}, the relatively lower resolution
of the KBOC timings makes anomaly detection more challenging, as higher
equal error rates have been associated with lower resolution timestamps.
Besides the global timer period, system load and other environmental
factors also play a role in determining the resolution and precision
of keystroke event timestamps \citep{killourhy2009role}.

\subsection{System evaluation}

Anomaly detectors are often evaluated by their equal error rate (EER),
the point on the receiver operating characteristic (ROC) curve at
which the false acceptance rate (FAR) is equal to the false rejection
rate (FRR). The EER can be derived by varying either a global threshold
or subject-dependent thresholds. 

To compute the \emph{global EER}, the ROC curve is derived by varying
a global threshold. In this case, the proportions of false positive
and false negative classifications from all subjects are counted to
compute the FAR and FRR at each threshold value.

The \emph{subject EER} is the EER obtained by deriving an ROC curve
for each subject. As a subject-dependent threshold is varied, the
FAR and FRR are computed using the number of false positive and false
negative classifications for a single subject. Since this produces
an EER for each subject in the dataset, system performance is typically
characterized by the mean and standard deviation (SD) of the distribution
of subject EERs. 

The KBOC anomaly detectors are evaluated by the global EER. There
are 20 query samples for each subject in the KBOC test dataset, which
provides $300\times20=6000$ scores to derive the global ROC curve.
In the rest of this paper, EER refers to the global EER unless otherwise
noted.

\section{Keystroke alignment \label{sec:Keystroke-alignment}}

\subsection{Motivation \label{sub:alignment-Motivation}}

The KBOC data collection procedure required samples to be a case-insensitive
match to the subject's first and last name. As a result, the presence
of modifier keys (e.g., \texttt{Shift} and \texttt{Caps Lock}) produced
different keystroke sequences for the same case-insensitive string
typed. Samples could vary in length (i.e., number of keystrokes) depending
on whether modifier keys were used for capitalization. They could
also vary in sequence depending on the order of keys pressed. In particular,
both template and query samples contain:
\begin{itemize}
\item \emph{Insertions and deletions}, e.g., due to the use or disuse of
\texttt{Shift}.
\item \emph{Substitutions}, e.g., due to the use of \texttt{Caps Lock} instead
of \texttt{Shift}\footnote{There are also some substitutions due to different keyboard scancode
sets being used \citep{scancodes}. In set 1, \texttt{Keypad-minus}
is 35 (hex), and in set 2 \texttt{Keypad-minus} is 4a (hex).}.
\item \emph{Transpositions}, e.g., both sequences ``\texttt{Space}, \texttt{Shift},
\texttt{T}'' and ``\texttt{Shift}, \texttt{Space}, \texttt{T}''
result in the string ``T'' when the \texttt{Shift} key is prolonged.
\end{itemize}
This scenario is not quite fixed text, such as password entry which
requires an exact match, and not quite free text, such as answering
an open-ended question which places no restriction on the text entered.
Instead, the keystroke sequence closely matches some target sequence
for each subject.

The need for a keystroke alignment method is verified by comparing
the keystroke sequences, and not the typed string. Consider all possible
combinations of template samples for each subject, i.e, ${4 \choose 2}\times300=1800$
combinations. Of these, 226 out of 1800 (12.6\%) template-to-template
comparisons differ from each other. The average Damerau-Levenshtein
(DL) distance, which measures insertions, deletions, substitutions,
and transpositions \citep{damerau1964technique}, is $0.053\pm0.276$
with a maximum distance of 4. Similarly, there are $20\times4\times300=24,000$
possible query-to-template combinations in which a query sample is
compared to a single template sample. Of these, 1545 out of 24,000
(6.4\%) query-to-template comparisons differ from each other, with
an average DL distance of $0.074\pm0.323$ and maximum distance of
7.

Many anomaly detectors operate on fixed-length feature vectors, motivating
the development of a method to compare slightly differing sequences.
One option is to use a free-text approach in which fixed-length feature
vectors are extracted from arbitrary keystroke sequences using a fallback
hierarchy of descriptive statistics \citep{monaco2013recent}. In
this case, the fact that the sequences are very close to some target
sequence is ignored. Instead, it would be desirable to create fixed-length
keystroke sequences from the semi-constrained sequences and then apply
standard feature extraction techniques \citep{killourhy2009comparing}.
The method introduced in this work aims to align the keystrokes from
two slightly differing sequences and then extract standard time interval
features from the aligned sequences.

\subsection{Alignment method \label{sub:alignment-Method}}

To compare slightly-differing keystroke sequences, a sequence-alignment
algorithm was developed to establish a correspondence between keystrokes
of two different samples. This allows two samples with similar keystroke
sequences to be compared using an anomaly detector that operates on
feature vectors of fixed length, such as Manhattan distance. In establishing
the correspondence, let the \emph{target} sequence be a fixed sequence
of keystrokes to which a \emph{given }sequence will be aligned. Thus,
the given sequence is modified to match the target.

\begin{figure}
\begin{centering}
\subfloat[Alignment.\label{fig:keystroke-align}]{\begin{centering}
\includegraphics[width=0.75\columnwidth]{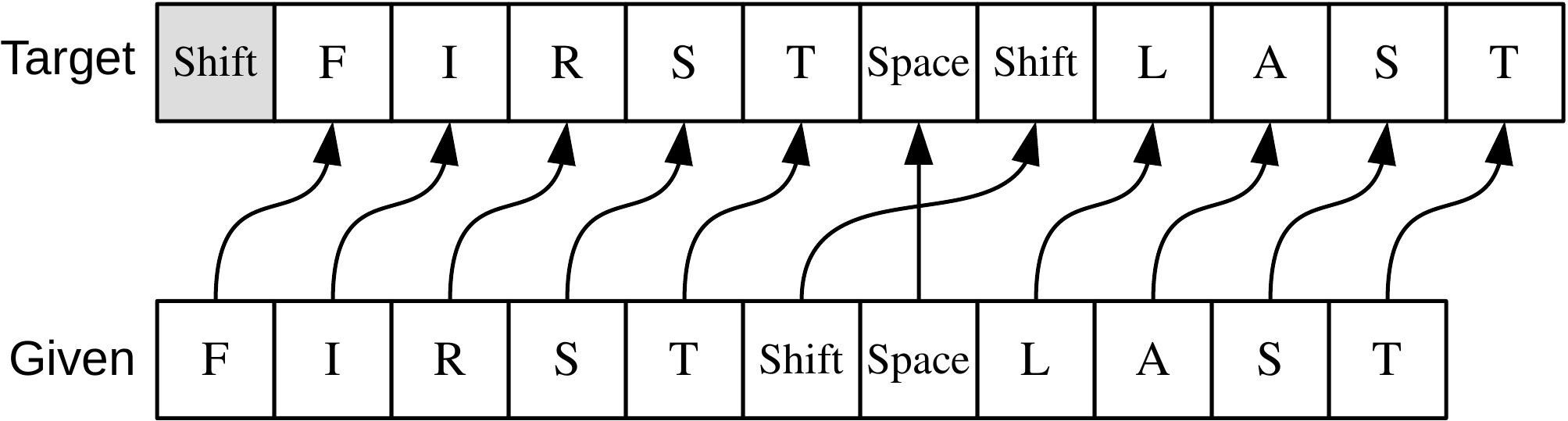}
\par\end{centering}

}
\par\end{centering}

\begin{centering}
\subfloat[Truncate.\label{fig:keystroke-truncate} ]{\begin{centering}
\includegraphics[width=0.75\columnwidth]{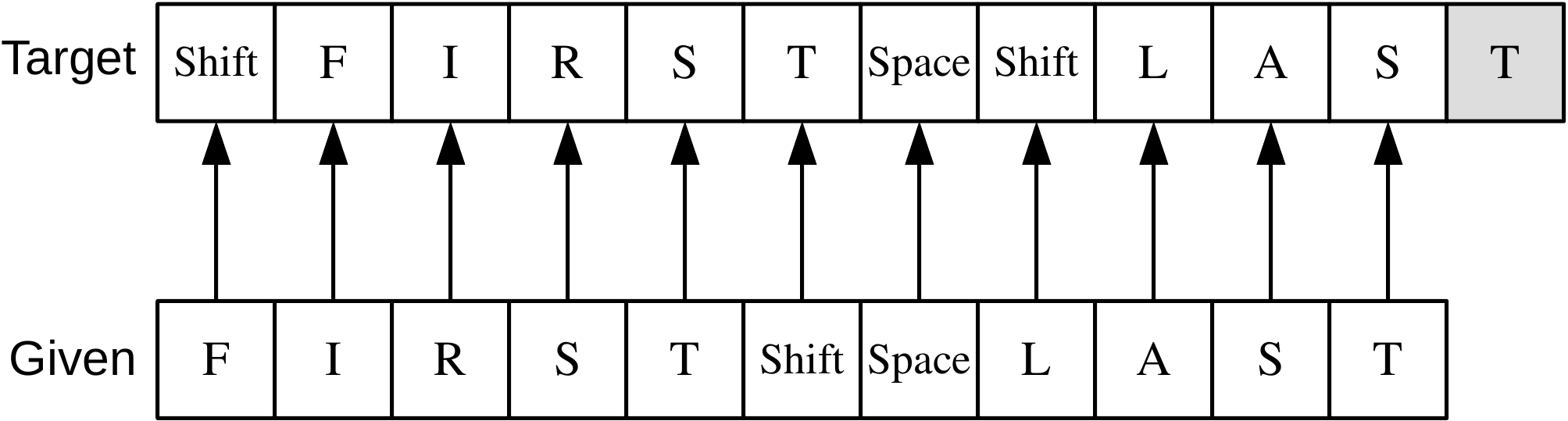}
\par\end{centering}

}
\par\end{centering}

\begin{centering}
\subfloat[Discard.\label{fig:keystroke-discard}]{\begin{centering}
\includegraphics[width=0.75\columnwidth]{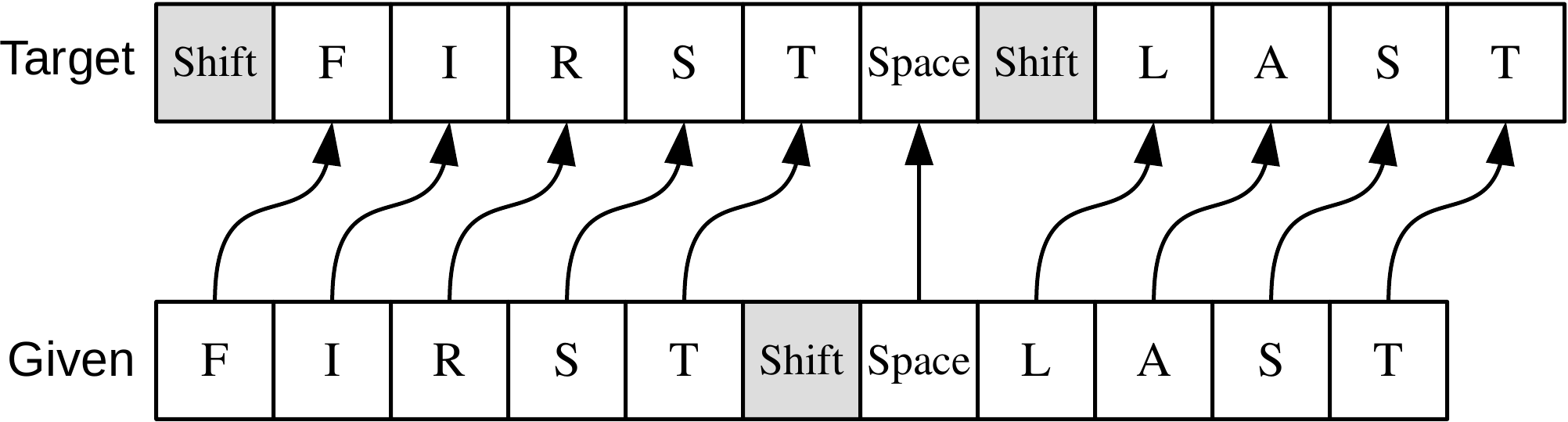}
\par\end{centering}

}
\par\end{centering}

\caption{Keystroke alignment vs.~two benchmark methods. In this example, the
first \texttt{Shift} is deleted in the given sequence, and ``\texttt{Shift},
\texttt{Space}'' are transposed. Keystroke alignment (\ref{fig:keystroke-align})
preserves the maximum amount of data, while ensuring a semantic correspondence
between keystrokes. Truncation (\ref{fig:keystroke-truncate}) ignores
key names, and is sensitive to sequences with differing lengths. Discarding
modifier keys (\ref{fig:keystroke-discard}) ensures keystrokes match
exactly (assuming case-insensitive equality), but ignores some of
the data. \label{fig:Keystrokes-align-methods}}
\end{figure}

Keystrokes from the given sequence are aligned to the target sequence
as follows. For each key in the target sequence, find the same key
in the given sequence with the closest position. If the key does not
exist in the given sequence, then use as a substitute the key in the
corresponding position. If a key appears in the given sequence, but
not the target sequence, it is simply ignored. The keystrokes in the
given sequence are then reordered by their mapping to the target sequence,
as shown in Figure \ref{fig:keystroke-align}.

Since there exist both within-template and query-to-template differences
in the keystroke sequences, as demonstrated in Section \ref{sub:alignment-Motivation},
a two-step alignment process is employed. First, let the target sequence
for each subject be the template sample with fewest keystrokes. The
remaining 3 template samples for each subject are aligned to the target
sequence. This creates a set of aligned template samples from which
the anomaly detector is trained. Next, the query samples for each
subject are aligned to the target sequence. The aligned query samples
are then scored by the fixed-text anomaly detector.

Alignment ensures an element-wise \emph{semantic similarity} between
the given and target keystrokes, and subsequently, the extracted features.
In other words, considering the keystrokes of a target sequence, the
corresponding keystrokes in the given sequence should have been pressed
with similar intention by the subject. Keystroke alignment ensures
this correspondence by matching substituted and transposed keys between
the two sequences.

To illustrate its utility, keystroke alignment is compared to some
benchmark alternatives, also shown in Figure \ref{fig:Keystrokes-align-methods}.
Let \emph{truncate} be the method in which the keystrokes of the given
sequence are simply truncated to be the same length as the target
sequence. This approach is shown in Figure \ref{fig:keystroke-truncate}.
Unless the keystroke sequences match exactly, semantically different
features will be compared by the anomaly detector, following feature
extraction. Truncation is especially sensitive to sequences that differ
in length, as shown in the example.

Alternatively, consider the method of \emph{discarding} modifier keys,
shown in Figure \ref{fig:keystroke-discard}. While this method has
the advantage that the resulting sequences will match exactly after
discarding modifier keys, it ignores distinguishable behavior which
may have been captured by modifier key usage. For example, if a subject
typically types ``\texttt{Left Shift}, \texttt{L}'' for a capital
``L'', and an impostor types ``\texttt{Right Shift}, \texttt{L}'',
the press-press latencies of the impostor will likely be longer than
that of the genuine subject. This is due keys that are far apart generally
being be pressed in quicker succession than keys that are close together
as a result of finger and/or hand reuse \citep{salthouse1984effects,monaco2015spoofing}.

\section{Feature extraction \label{sec:Feature-extraction}}

Time interval features are extracted from the aligned keystroke sequence,
wherein each keystroke event contains a press time, release time,
and key name. While a variety of keystroke timings can be computed
from the timestamps, this work uses the press-press latencies and
key-hold durations. The press-press latency is the time interval between
two successive key-presses, denoted by $p$. The key-hold duration
is the time interval from the press to release of each key, denoted
by $d$. If the target sequence contains 11 keystrokes, then the feature
vectors for every given sequence aligned to the target will contain
31 features (10 latencies and 11 durations). Note that any other time
interval, such as release-press or release-release latencies of successive
keystrokes, can be formed by a linear combination of the press-press
latency and key-hold duration.

\begin{center}
\begin{figure}
\begin{centering}
\includegraphics[width=1\columnwidth]{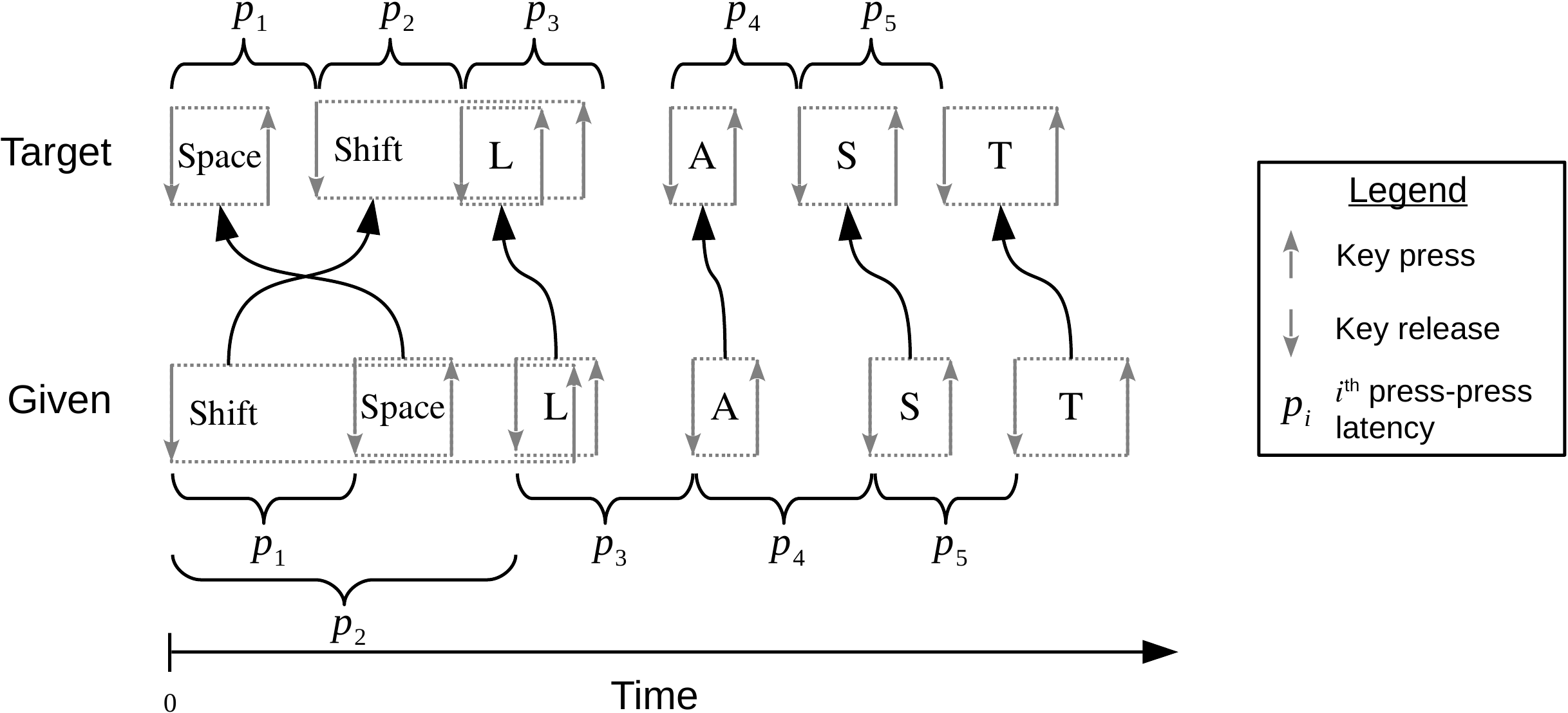}
\par\end{centering}

\caption{Keystroke features.\label{fig:keystroke-features}}
\end{figure}

\par\end{center}

The press-press latency is typically positive since keystrokes are
normally ordered by their press time (as they would appear on screen
to the user). After a given sequence has been aligned to a target
sequence, the press-press latency can be negative when there is a
transposition between the given and target sequence. Using Figure
\ref{fig:keystroke-features} as an example, the keystrokes in the
given sequence are reordered to the positions of the corresponding
keys in the target sequence. In the given sequence, the press-press
latency from \texttt{Space} to \texttt{Shift}, denoted by $p_{1}$,
is negative since \texttt{Shift} is pressed before \texttt{Space}.

Reordering the keystrokes of the given sequence in this way, such
that the press-press latency becomes negative for transposed keys,
does not have an adverse effect on anomaly detection as long as this
behavior is also captured in the template samples. Consider a typist
who presses the keys \texttt{Space} and \texttt{Shift} in very quick
succession. The press-press latency between these keys will be close
to 0, and if normal behavior is to transpose these keys 50\% of the
time, the distribution of press-press latencies will be centered on
0. Then, it does not matter whether the feature in the given sequence
is positive or negative, i.e., whether the given sequence contains
``\texttt{Space}, \texttt{Shift}'' or ``\texttt{Shift}, \texttt{Space}'';
it has equal likelihood in both cases.

\subsection{Feature normalization \label{sub:Feature-normalization}}

The latency and duration features are normalized to the range $[0,1]$.
The $i$th normalized press-press latency $\dot{p}_{i}$ is given
by

\begin{equation}
\dot{p}_{i}=\max\left[0,\min\left(1,\frac{p_{i}-\lfloor p\rfloor}{\lceil p\rceil-\lfloor p\rfloor}\right)\right]\label{eq:latency-normalization}
\end{equation}
where $p_{i}$ is the raw feature and $\lfloor p\rfloor$ and $\lceil p\rceil$
provide the lower and upper bounds for normalization, respectively.
Let $\mu_{p}$ be the mean and $\sigma_{p}$ the SD of the distribution
of latencies in the subject's aligned template samples. That is, $\mu_{p}$
and $\sigma_{p}$ are determined by the aligned templates no matter
whether the feature being normalized is from a query or template sample.
Normalization bounds are given by 
\begin{eqnarray}
\lfloor p\rfloor & = & \mu_{p}-H_{f}\sigma_{p}\nonumber \\
\lceil p\rceil & = & \mu_{p}+H_{f}\sigma_{p}\label{eq:feature-normalization}
\end{eqnarray}
where $H_{f}$ is a free parameter. In this work, results were obtained
using $H_{f}=1$, i.e., normalizing features to within one SD of the
mean. The duration features are normalized similarly, replacing $d$
for $p$. 

Note that feature normalization uses the first and second order statistics
of the distribution of the subject's aligned template samples. The
effect of this approach is twofold. First, it ensures that outlier
features do not greatly affect the normalization bounds as they would
using min/max normalization. Second, per-subject normalization ensures
that normalized features are relative to the subject's template since
each subject has their own normalization bounds. Query samples are
normalized using the lower and upper bounds determined from the template
samples. Thus, if the query sample is genuine, its normalized features
will be closer to the center of the interval $[0,1]$, whereas the
normalized features of an impostor sample will be closer to the endpoints
(assuming it is greater or less than the genuine mean).

\section{Anomaly detection \label{sec:Anomaly-detection}}

The keystroke alignment and feature normalization methods introduced
in this work are agnostic to the particular anomaly detector used.
Six different anomaly detectors, and one ensemble method, are described
in this section. Neural network models were implemented using the
TensorFlow library \citep{abadi2016tensorflow}. All models were implemented
in Python and remain publicly available at \url{https://github.com/vmonaco/kboc}.

\subsection{Autoencoder}

The basic autoencoder (AE) is a neural network that aims to encode
and then decode its input \citep{bengio2009learning}. The network
topology consists of equally-sized input and output layers separated
by at least one hidden layer. Successive layers are fully connected
through weight matrices, a bias vector, and a nonlinear function.
It is common to use tied weights, i.e., the weight matrix connecting
the hidden and output layers is simply the transpose weight matrix
connecting the input and hidden layers. This limits the number of
model parameters and acts as a kind of regularization. 

Using feature vector $x$ as input, the hidden layer is calculated
as
\begin{equation}
h=f\left(x\right)=\tanh\left(Wx+b_{h}\right)\label{eq:ae-hidden}
\end{equation}
where $W$ is the weight matrix and $b_{h}$ is the hidden layer bias
vector. Similarly, the output layer $y$ is given by
\begin{equation}
y=g\left(x\right)=\tanh\left(W^{\top}h+b_{y}\right)\label{eq:ae-output}
\end{equation}
where $b_{y}$ is the output bias vector. The complete model parameters
consist of $\theta=\left\{ W,b_{h},b_{y}\right\} $. Hidden layers
can be stacked to create a deep network, and dimensions of the hidden
layers comprise the model hyperparameters. Typically, at least one
hidden layer should be smaller than the input layer to achieve a compressed
representation and avoid learning the identity function. Parameters
are determined by back-propagating gradients from a squared error
loss function,

\begin{equation}
L\left(x,y\right)=\|x-y\|^{2}\;.\label{eq:ae-loss}
\end{equation}
where $y$ is the reconstructed output. The objective function of
the basic autoencoder is given by
\begin{equation}
\mathcal{J}_{AE}\left(\theta\right)=\sum_{x\in D}L\left(x,g\left(f\left(x\right)\right)\right)\label{eq:ae-objective}
\end{equation}
where $D$ is the set of training examples. Bias parameters are initialized
to 0 and weights are initialized from a random uniform distribution
such that the scale of the gradients in each layer is roughly the
same \citep{glorot2010understanding}. During testing, the score of
a query sample is given by the negative reconstruction error, $-L\left(x,y\right)$.

\subsection{Contractive autoencoder}

The contractive autoencoder (CAE) is an autoencoder that uses the
Frobenius norm of the hidden layer Jacobian as a regularization term
\citep{rifai2011contractive}. This enables a sparse representation
of the input whereby the dimension of the hidden layer is much larger
than the input and output layer dimensions. The CAE is closely related
to the denoising autoencoder, in which the goal is to reconstruct
an input vector that has been corrupted by noise \citep{vincent2008extracting}.

The CAE in this work uses a sigmoid activation function. The hidden
layer is given by

\begin{equation}
h=f\left(x\right)=\sigma\left(Wx+b_{h}\right)\label{eq:cae-hidden}
\end{equation}
where $\sigma\left(z\right)=\left(1+e^{-z}\right)^{-1}$. Similarly,
the output layer is given by 
\begin{equation}
y=g\left(x\right)=\sigma\left(W^{\top}h+b_{y}\right)\;.\label{eq:cae-output}
\end{equation}
The objective function of the CAE includes a regularization term that
penalizes the Jacobian of $f$,

\[
\mathcal{J}_{CAE}\left(\theta\right)=\sum_{x\in D}L\left(x,g\left(f\left(x\right)\right)\right)+\lambda\|J_{f}\left(x\right)\|_{F}^{2}
\]
where $\lambda$ is a free parameter and $\|\cdot\|_{F}^{2}$ is the
squared Frobenius norm. Similar to the basic autoencoder, the loss
function, $L$, is the squared reconstruction error. Using a sigmoid
activation function allows for efficient computation of the Jacobian,
given by

\begin{equation}
\|J_{f}\left(x\right)\|_{F}^{2}=\sum_{ij}\left[\frac{\partial h_{j}\left(x\right)}{\partial x_{i}}\right]^{2}=\sum_{i=1}^{d_{h}}\left[h_{i}\left(1-h_{i}\right)\right]^{2}\sum_{j=1}^{d_{x}}W_{ij}^{2}\label{eq:sigmoid-jacobian}
\end{equation}
where $d_{h}$ and $d_{x}$ are the hidden and input layer dimensions,
respectively. Similar to the basic AE, bias vectors are initialized
to 0 and weight vectors use Xavier initialization \citep{glorot2010understanding}.

\subsection{Variational autoencoder}

The variational autoencoder (VAE) is a probabilistic autoencoder with
continuous latent variables \citep{kingma2013auto}. Parameters are
learned efficiently by backpropagation through a \emph{reparametrization
trick}, which allows the gradient of the loss function to propagate
through the sampling process. The objective function of the VAE is
composed of both a reconstruction loss and a latent loss. Query sample
scores are given by the negative reconstruction loss, which is the
negative log probability of the input given the reconstructed latent
distribution.

\subsection{One-class support vector machine}

The one-class support vector machine (SVM) is an unsupervised model
that learns a separating hyperplane between the origin and feature
vector points\footnote{Alternatively, one may learn a minimal-volume hypersphere that encapsulates
most of the points in feature space.} \citep{scholkopf2001estimating}. The free parameters of the model
include $\eta$, the fraction of training errors (samples that lie
outside the separating plane), in addition to any parameters of the
kernel function. This work uses the one-class SVM implemented by scikit-learn
\citep{pedregosa2011scikit}, which internally uses libsvm \citep{chang2011libsvm}.
Query sample scores are given by the negative distance to the separating
hyperplane.

\subsection{Manhattan distance}

Manhattan distance between two vectors $x$ and $y$ is computed by
\begin{equation}
D\left(x,y\right)=\sum_{i}^{d_{x}}|x_{i}-y_{i}|\label{eq:manhattan-dist}
\end{equation}
where $d_{x}$ is the dimension of the feature vectors. Query sample
scores are given by the negative Manhattan distance to the mean template
vector. Thus, training is comprised simply of computing the element-wise
mean of the template vectors.

Scaled Manhattan distance was previously shown to achieve state-of-the-art
accuracy in keystroke biometric anomaly detection \citep{killourhy2009comparing},
and Manhattan distance closely followed. Scaled Manhattan distance
was not used in this work since the number of template samples (4)
is not large enough to obtain a reliable estimate of the absolute
(or standard) deviation of each feature.

\subsection{Partially observable hidden Markov model}

The partially observable hidden Markov model (POHMM) is an extension
of the hidden Markov model in which hidden states are conditioned
on an independent Markov chain \citep{monaco2015time,monaco2016partially}.
In a two-state model of typing behavior, the subject can be in an
active state, during which relatively short time intervals are observed,
or passive state, during which relatively long time intervals are
observed. The keyboard key names form an independent Markov chain
upon which the hidden states are conditioned. The key name partially
reveals the hidden state since certain keys, such as \texttt{Space}
or \texttt{Shift}, indicate a greater probably of being in a passive
state, i.e., of observing longer time intervals, than letter keys,
such as \texttt{H} and \texttt{E}. Despite an explosion in the number
of model parameters, parameter estimation can still be performed in
linear time using a modified Baum-Welch algorithm. A parameter smoothing
technique is employed to avoid overfitting the model to short input
sequences \citep{monaco2016partially}.

Keystroke timing features are modeled by a lognormal distribution,
and parameters are initialized based on a one-state model. The POHMM
is a free-text model, since it operates on arbitrary keystroke sequences
and does not require fixed-length input. Therefore, the POHMM does
not use the keystroke alignment or feature extraction algorithms described
in Sections \ref{sec:Keystroke-alignment} and \ref{sec:Feature-extraction},
respectively. Query sample scores are given by the model loglikelihood.

\subsection{Ensemble}

An ensemble of anomaly detectors was formed with the goal of achieving
higher accuracy than any individual member. The ensemble score is
simply the mean unnormalized score from each system in the ensemble.
The sum rule (which is equivalent to a scaled mean) has been shown
to be a robust, albeit simple, method of combining classifier output
scores, especially when the scores between the systems being combined
are independent \citep{kittler1998combining}.

\section{Score normalization \label{sec:Score-normalization}}

Per-subject score normalization is employed whereby normalization
bounds are determined by the test scores for each subject. Let $s_{u}$
be the set of scores from subject $u$ and $s_{ui}$ be the $i$th
score from subject $u$. The normalized score, $\dot{s}_{ui}$, is
given by

\begin{equation}
\dot{s}_{ui}=\max\left[0,\min\left(1,\frac{s_{ui}-\lfloor s_{u}\rfloor}{\lceil s_{u}\rceil-\lfloor s_{u}\rfloor}\right)\right]\label{eq:normalization}
\end{equation}
where $\lfloor s_{u}\rfloor$ and $\lceil s_{u}\rceil$ are the lower
and upper bounds used for normalization. Note that the clamping function
$\max\left[0,\min\left(1,\cdot\right)\right]$ ensures the resulting
score is in the range $\left[0,1\right]$. Two methods of score normalization
are compared, which differ in the way $\lfloor s_{u}\rfloor$ and
$\lceil s_{u}\rceil$ are defined.

\subsection{SD normalization}

In \emph{SD normalization}, the scores are normalized using the SD
of the scores within each subject. Let $\mu_{s_{u}}$ be the mean
score for subject $u$ and $\sigma_{s_{u}}$ be the SD score for subject
$u$. The lower and upper bounds for normalization are given by
\begin{eqnarray}
\lfloor s_{u}\rfloor & = & \mu_{s_{u}}-H_{s}\sigma_{s_{u}}\nonumber \\
\lceil s_{u}\rceil & = & \mu_{s_{u}}+H_{s}\sigma_{s_{u}}\label{eq:sd-normalization}
\end{eqnarray}
where $H_{s}$ is a free parameter. In this work, $H_{s}=2$ normalizes
scores to within 2 SD of the mean.

\subsection{Min/max normalization \label{sub:Min/max-normalization}}

Min/max normalization is achieved by letting

\begin{eqnarray}
\lfloor s_{u}\rfloor & = & \min s_{u}\nonumber \\
\lceil s_{u}\rceil & = & \max s_{u}\label{eq:min-max-normalization}
\end{eqnarray}
where $\min s_{u}$ and $\max s_{u}$ denote the minimum and maximum
scores from subject $u$, respectively.

\section{Results\label{sec:Results}}

\begin{table}
\begin{centering}
\begin{tabular}{|c|l|r@{\extracolsep{0pt}.}l|r@{\extracolsep{0pt}.}l|}
\hline 
System & Description & \multicolumn{2}{c|}{Validation} & \multicolumn{2}{c|}{Test}\tabularnewline
\hline 
\hline 
\multicolumn{6}{|c|}{SD score normalization}\tabularnewline
\hline 
\hline 
1 & Autoencoder \{5,4,3\} & 7&90 (1.91) & 7&82\tabularnewline
\hline 
2 & Variational autoencoder & 6&50 (0.75) & 6&46\tabularnewline
\hline 
3 & POHMM & 7&75 (1.36) & 7&32\tabularnewline
\hline 
4 & One-class SVM & 7&40 (2.20) & 7&35\tabularnewline
\hline 
5 & Contractive autoencoder \{400\} & 6&20 (1.27) & 8&02\tabularnewline
\hline 
6 & Manhattan distance & 6&95 (1.17) & \textbf{5}&\textbf{32}\tabularnewline
\hline 
7 & Autoencoder \{5\} & 5&75 (1.57) & 7&95\tabularnewline
\hline 
8 & Contractive autoencoder \{200\} & 6&50 (1.47) & 8&08\tabularnewline
\hline 
9 & Mean ensemble systems 3,4,5 & 5&40 (1.45) & 5&68\tabularnewline
\hline 
10 & Mean ensemble systems 1-8 & 5&85 (1.31) & 5&91\tabularnewline
\hline 
\hline 
\multicolumn{6}{|c|}{Min/max score normalization}\tabularnewline
\hline 
\hline 
11 & POHMM & 8&75 (1.38) & 10&35\tabularnewline
\hline 
12 & One-class SVM  & 12&10 (1.63) & 10&89\tabularnewline
\hline 
13 & Contractive autoencoder \{400\} & 7&85 (1.23) & 11&20\tabularnewline
\hline 
14 & Contractive autoencoder \{200\} & 8&00 (1.25) & 11&23\tabularnewline
\hline 
15 & Mean ensemble systems 11-14 & 5&75 (1.30) & 6&26\tabularnewline
\hline 
\end{tabular}
\par\end{centering}

\caption{Summary of validation and test EER (\%). The validation EER was obtained
by Monte Carlo procedure that mimics test set conditions (SD shown
in parentheses). The test EER was determined by the KBOC organizers
using score submission files. Hidden layer sizes of neural network
models are shown in braces.\label{tab:EER-summary}}
\end{table}

Hyperparameters, such as hidden layer sizes in the AE models and convergence
criteria, were selected from a small range that performed well on
the KBOC development set. A validation EER was determined for each
system through a Monte Carlo validation procedure using the development
dataset. In each repetition, 4 template samples were randomly selected
from the genuine samples. The remaining 20 samples (10 genuine and
10 impostor) were used as the query samples. These conditions mimic
the test dataset, in which there are 4 labeled template samples and
20 unlabeled query samples for each subject. This process was repeated
10 times to obtain a confidence interval on the EER. System descriptions,
including hyper-parameters, are as follows:
\begin{description}
\item [{System~1:}] AE with hidden layers of dimensions \{5, 4, 3\}. Models
were trained for 5000 epochs using gradient descent with a learning
rate of 0.5.
\item [{System~2:}] VAE with hidden layers of dimensions \{5, 5\}, 3-dimensional
Gaussian latent space, and softplus nonlinearities between layers.
Parameters were learned by \emph{Adam}, a stochastic gradient-based
optimization algorithm \citep{kingma2014adam}, using a learning rate
of 0.001, mini-batch size of 2, and 700 epochs.
\item [{Systems~3~and~11:}] POHMM with two-dimensional log-normal emission
(for the press-press latency and duration) and two hidden states.
Convergence was reached when the reduction in the loglikelihood was
less than 0.01.
\item [{Systems~4~and~12:}] one-class SVM with $\eta=0.5$, radial basis
function (RBF) kernel, and RBF kernel parameter $\gamma=0.9$.
\item [{Systems~5~and~13:}] CAE with hidden layer of dimension 400 and
regularization weight $\lambda=1.5$. Model parameters were learned
by gradient descent with a 0.01 learning rate and 1000 epochs. 
\item [{System~6:}] Manhattan distance.
\item [{System~7:}] AE with hidden layer of dimension 5.
\item [{Systems~8~and~14:}] CAE with hidden layer of dimension 200 and
regularization weight $\lambda=0.5$.
\item [{System~9:}] Ensemble mean score of systems 3, 4, and 5, which
have low score correlations. The correlation between scores of systems
3 and 4 is 0.665, the lowest of any pair that utilized SD score normalization. 
\item [{System~10:}] Ensemble mean score of systems 1-8, which all use
SD score normalization.
\item [{System~15:}] Ensemble mean score of systems 11-14, which use min/max
score normalization.
\end{description}
Additionally, systems 1-10 used SD score normalization, and systems
11-15 used min/max score normalization.

The resulting validation and test EERs of the fifteen systems submitted
to the KBOC are shown in Table \ref{tab:EER-summary}. The test EERs
were determined by the KBOC organizers, as the ground truth labels
of the query samples remained hidden from competition participants.
Systems 1-10, which used SD score normalization, generally obtained
lower EERs than systems 11-15, which used min/max score normalization.
Systems 9 and 15, both ensembles, achieved a lower EER on the test
set than any individual member in the ensemble. This is not the case
for system 10, which is an ensemble of systems 1-8. Systems 5, 7,
and 8 show a significant decrease in performance from the validation
EER, suggesting overfitting to the development set due to the chosen
hyperparameters.

\section{Discussion\label{sec:Discussion}}

\subsection{Effects of keystroke alignment}

\citet{bours2014performance} outlined three different scenarios that
can occur as a result of typing fixed text, such as password entry:
\begin{enumerate}
\item The string that was typed does not match the target string.
\item The string that was typed matches the target string, but contains
corrections due to typing errors.
\item The string that was typed matches the target string without making
corrections. This is the most commonly assumed scenario in keystroke
biometrics, such that the query keystroke sequences exactly match
the template sequence.
\end{enumerate}
\citet{bours2014performance} considered scenario 2, in which sequences
could contain extra keystrokes, such as \texttt{Backspace}, due to
typing errors. They concluded that system performance generally increased
as a result of lower FTC and more data being available. However, their
work did not utilize keystrokes outside of the target sequence (\texttt{.tie5Roanl}),
as the data was simulated from the fixed-text dataset in \citet{killourhy2009comparing}
which assumed scenario 3. The KBOC dataset instead reflects scenario
1, in which the string that was typed approximately matches the target
string.

In order to further reduce the FTC rate from scenario 2, typed strings
that closely match, but slightly differ from, the target string may
be retained (scenario 1). The KBOC dataset reflects this, as only
case-insensitive matches to the target string were required. This
resulted in differing keystroke sequences, and to handle these differences,
a keystroke alignment technique was developed. To determine the effect
of keystroke alignment, the method proposed in this work is compared
to the benchmark alternatives described in Section \ref{sub:alignment-Method}:
\begin{description}
\item [{Discard:}] Discard modifier keys, comparing only character keys.
\item [{Truncate:}] Truncate sequences to the length of the shortest sample.
\end{description}
Table \ref{tab:effects-alignment} shows the effects of keystroke
alignment using a Manhattan distance detector and SD score normalization.
The best performance is achieved using the keystroke alignment method.
Discarding modifier keys obtains the worst performance, suggesting
that the use of modifier keys plays an important role in keystroke
biometrics. Truncating sequences to be of equal length results in
a slightly higher EER than keystroke alignment. Truncation causes
different features to be compared to each other for sequences of differing
length. For example, if the template begins with \texttt{Shift}, and
the query begins with \texttt{T}, the \texttt{Shift} duration feature
will correspond to the \texttt{T} duration feature. This has an adverse
effect if the query sample is genuine. However, if the query sample
is an impostor, the mismatched features may generally have greater
distances than the aligned features, improving overall performance.
Note that the validation EERs of the truncation and alignment methods
are equal since the length of the samples in the development set did
not differ within any of the 10 subjects.

\begin{table}
\begin{centering}
\begin{tabular}{|c|r@{\extracolsep{0pt}.}l|r@{\extracolsep{0pt}.}l|r@{\extracolsep{0pt}.}l|}
\cline{2-7} 
\multicolumn{1}{c|}{} & \multicolumn{2}{c|}{Discard} & \multicolumn{2}{c|}{Truncate} & \multicolumn{2}{c|}{Align}\tabularnewline
\hline 
Validation & 10&05~(1.44) & 6&95~(1.17) & 6&95~(1.17)\tabularnewline
\hline 
Test & 9&18 & 7&33 & 6&53\tabularnewline
\hline 
\end{tabular}
\par\end{centering}

\caption{EER (\%, SD in parentheses) obtained using different keystroke alignment
methods with a Manhattan detector and SD score normalization. Discard=discard
modifier keys, Truncate=truncate the template and query keystroke
sequences to the shortest template, Align=keystroke alignment method
described in Section \ref{sec:Keystroke-alignment}. \label{tab:effects-alignment}}
\end{table}

\subsection{Effects of score normalization}

\begin{figure}
\begin{centering}
\includegraphics[width=1\textwidth]{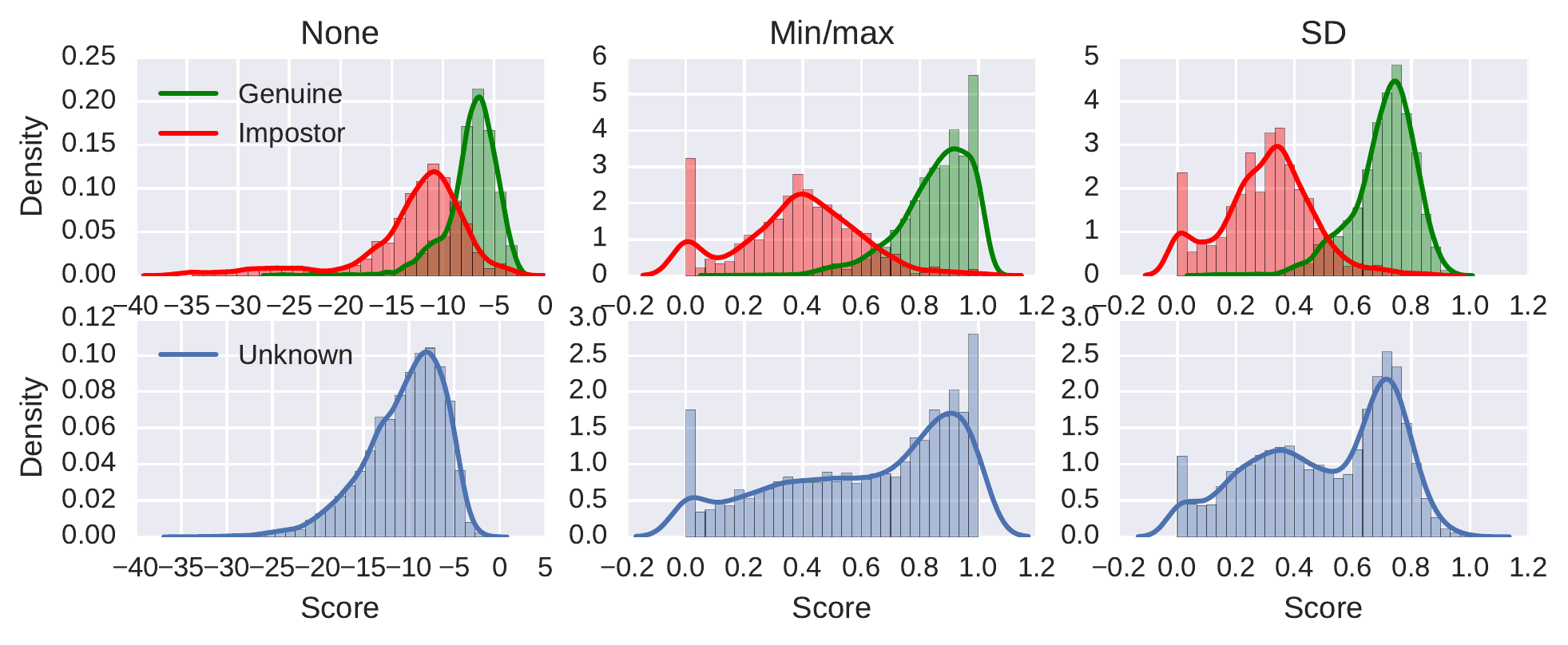}
\par\end{centering}

\caption{The effects of score normalization using a Manhattan distance anomaly
detector: unnormalized (left), min/max normalization (middle), SD
normalization (right). Min/max normalization is sensitive to outliers;
SD normalization is not.\label{fig:score-distributions}}
\end{figure}

To determine the effects of score normalization, the EER is determined
using each normalization method with a Manhattan detector and keystroke
alignment. Table \ref{tab:effects-score-normalization} contains the
results, and Figure \ref{fig:score-distributions} shows the resulting
score distributions. The unnormalized scores perform significantly
worse than either min/max or SD normalization, while SD normalization
achieves the best performance. The unnormalized scores of the test
set appear to have a unimodal distribution. The min/max normalized
scores appear bimodal at 0 and 1 due to the presence of 0 and 1 scores
for each subject. This effect is not present in the SD normalized
scores, which appear to have a trimodal distribution with peaks at
0 and above the (presumably) expected genuine and impostor scores.
Score normalization can have a significant effect on system performance
and remains an ongoing area of research \citep{moralesscore}. System
performance, as measured by either the global or subject EER, is affected
by score normalization in three different ways.

\begin{table}
\begin{centering}
\begin{tabular}{|c|r@{\extracolsep{0pt}.}l|r@{\extracolsep{0pt}.}l|r@{\extracolsep{0pt}.}l|}
\cline{2-7} 
\multicolumn{1}{c|}{} & \multicolumn{2}{c|}{None} & \multicolumn{2}{c|}{Min/max} & \multicolumn{2}{c|}{SD}\tabularnewline
\hline 
Validation & 18&40~(1.90) & 7&55~(0.96) & 6&95~(1.17)\tabularnewline
\hline 
Test & 21&20 & 10&43 & 6&53\tabularnewline
\hline 
\end{tabular}
\par\end{centering}

\caption{EER (\%, SD in parentheses) obtained using different score normalization
methods with a Manhattan detector and keystroke alignment. \label{tab:effects-score-normalization}}
\end{table}

First, score normalization places the scores from multiple anomaly
detectors on the same scale for the purpose of score-level classifier
fusion \citep{jain2005score}. Members of an ensemble may correspond
to different models for a single modality or to different modalities.
Normalization ensures that the scores for each anomaly detector are
equally weighted or combined in a way that reflects the relative performance
of each underlying model.

Second, score normalization places the scores from different subjects
on the same scale. This is only relevant for performance metrics derived
from a single global threshold, such as the global ROC curve and global
EER. As the score distributions of different subjects vary, a single
global threshold may correspond to vastly different FAR and FRR for
each subject. To achieve similar FAR and FRR for each subject with
a global threshold, the subject scores must be normalized. Score normalization,
in this sense, does not greatly affect the subject EER which is based
on subject-specific thresholds that vary independently. Specifically,
the min/max score normalization described in Section \ref{sub:Min/max-normalization}
results in exactly the same subject EER since the shape of the score
distribution for each subject is preserved.

Finally, score normalization can reshape the score distribution within
a single subject. This can have the effect of being robust to sample
outliers and lead to a better separation of genuine and impostor scores.
Like tanh normalization \citep{jain2005score}, the SD score normalization
described in this work is robust to sample outliers, whereas min/max
score normalization is sensitive to outliers.

\section{Conclusions and future work\label{sec:Conclusions}}

The time between data collection sessions for each subject ranged
from less than a day to several months, with an average interval of
1 month. This captured a wide range of variability in typing behavior.
Despite this, the best system developed in this work (system 6) remained
robust over long intervals, as noted in \citet{morales2016keystroke},
having the lowest decrease in performance when testing query samples
collected 2 months after the template (5.09\% EER) compared to those
collected 4 months after the template (5.10\% EER). The relatively
small decrease in performance of all systems evaluated in the KBOC
(the greatest being a 7\% relative increase \citep{morales2016keystroke})
may also be attributed to the stabilization of subjects' templates.
Keystroke sequences have been shown to converge towards a \emph{total
profile}, i.e., a stationary feature distribution, with the number
of typing repetitions \citep{montalvao2015contributions}. First and
last names are likely practiced many times by the genuine subjects,
each template having reached the subject's total profile.

Manhattan distance is arguably the simplest anomaly detector out of
the fifteen systems evaluated in this work since it has no free parameters,
yet it achieved the lowest EER on the test dataset of 5.32\%. Since
feature normalization scales the features by their SD, Manhattan distance
in this work is actually similar to the scaled Manhattan distance
on the raw features. Taking the best submission (system 6) and using
min/max feature normalization yields a validation EER of 8.55 (1.21)\%
and test EER of 8.60\%. This result is consistent with \citet{killourhy2009comparing},
wherein scaled Manhattan obtained the best performance. Since the
publication of \citet{killourhy2009comparing}, other anomaly detectors,
such as Gaussian mixture models \citep{deng2013keystroke,deng2015keystroke}\footnote{The deep neural network and other methods in \citet{deng2013keystroke,deng2015keystroke}
utilize negative data for training.}, have achieved only marginally-better performance than the scaled
Manhattan distance detector. The relative worse performance of other
systems in this work over the Manhattan detector indicates overfitting
to the template samples, while an increase in test EER from validation
EER indicates overfitting to the development set due to hyperparameter
selection.

Unlike other keystroke biometric datasets, such as \citet{killourhy2009comparing},
the input sequences (first and last name) were unique to each subject
in the KBOC dataset. Therefore, only anomaly detection approaches
were investigated in this work, as negative data could not practically
be utilized for training a discriminative model. Future work could
examine the use of negative training data when keystroke sequences
are unique to each subject and highly practiced, as is the case in
the KBOC dataset.

Lastly, like most previous works, the results is this work reflect
asymptotic system performance and may be considered forensics rather
than biometrics. Both the min/max and SD score normalization methods
utilize statistics of the empirical query score distribution, which
requires having observed many query samples. Future work should investigate
online score normalization, which is subject to the order in which
genuine and impostor samples appear to the system.

\section*{References}

\bibliographystyle{plainnat}
\bibliography{robust-keystroke}

\end{document}